\documentclass[english,conference]{IEEEtran}
\usepackage[T1]{fontenc}
\usepackage[latin9]{inputenc}
\usepackage{url}
\usepackage{amsthm}
\usepackage{amsmath}
\usepackage{graphicx}
\usepackage{amssymb}

\makeatletter
\theoremstyle{plain}
\newtheorem{thm}{Theorem}
  \theoremstyle{plain}
  \newtheorem{lem}{Lemma}
   \theoremstyle{remark}
  \newtheorem{rem}{Remark}

\makeatother

\usepackage{babel}

\begin{document}

\title{The Capacity of a Class of Linear Deterministic Networks }

\author{\IEEEauthorblockN{S. M. Hossein Tabatabaei Yazdi and Mohammad Reza Aref}
 \IEEEauthorblockA{Information Systems and Security Lab (ISSL)\\
 Department
of Electrical Engineering,
 Sharif University of Technology,
 Tehran, Iran\\
E-mail: \url{smh_tabatabaei@ee.sharif.edu} and \url{aref@sharif.edu}} }
\maketitle
\footnote{This work was partially supported by Iranian National Science Foundation (INSF) under contact No 84,5193-2006. }\begin{abstract}
In this paper, we investigate optimal coding strategies for a 
class of linear deterministic relay networks. The network under study
is a relay network, with one source, one destination, and two relay
nodes. Additionally, there is a disturbing source of signals that
causes interference with the information signals received by the relay
nodes. Our model captures the effect of the interference of message
signals and disturbing signals on a single relay network, or the interference
of signals from multiple relay networks with each other in the linear
deterministic framework. For several ranges of the network parameters
we find upper bounds on the maximum achievable source--destination
rate in the presense of the disturbing node and in each case we find
an optimal coding scheme that achieves the upper bound.
\end{abstract}
\IEEEpeerreviewmaketitle

\section{Introduction}
Finding the capacity of relay networks is a basic open problem in network information theory \cite{cover}. One way to approximate the function of relay networks is through deterministic models that was first introduced by Aref \cite{aref}.    
 Avestimehr, Diggavi and Tse have recently \cite{avestimehr_1,avestimehr_2}
introduced a linear deterministic model for wireless relay networks that
treats noise as a deterministic thresholding function and the interference
of signals as a linear transformation over a finite field. They have
successfully applied this model to several relay networks and have
shown that the capacity of the deterministic model is within a constant
gap from the corresponding wireless network. Furthermore, they have
shown that a max--flow min--cut result holds for the capacity of the
relay deterministic network in the case of a single multicast session.
While the capacity achieving scheme in \cite{avestimehr_1,avestimehr_2}
is a random coding over long blocks of signals, recent works \cite{fragouli_1,goemans,tabatabaei_yazdi_soda,tabatabaei_yazdi_itw2010}
have devised low complexity and deterministic schemes that achieves
the maximum capacity in the case of a unicast session. 

In the case of multiple messages, Mohajer et al., \cite{mohajer_diggavi_fragouli}
have considered two unicast sessions on a relay network with two relays,
two sources, and two destinations. They study two special cases of
this setting, namely ZS and ZZ channels and give a full characterization
of the capacity region along with capacity acheiving schemes in each
case. 

In this paper, we are also interested in the deterministic relay network
with interference at each node. The network here consists of a source
node, a destination node, and two relay nodes. Additionally there
is a disturbing node that sends signals to the two relays and causes
interference with message signals. Our goal is to characterize the
capacity region from source to destination in the presense of the
disturbing node. Our model differs from the model in \cite{mohajer_diggavi_fragouli}
in two ways. First they consider a two dimensional capacity region
in which each source node tryes to send messages to its designated
destination. But in our model, we only try to find the one dimensional
capacity from one source to the corresponding destination. On the
other hand, while in both the ZS and ZZ channels the interference
of the two messages only occur in one relay node, in our model we assume
that the disturbing signals have interference with both relays which
makes it more difficult to handle. We find the capacity
and offer capacity achieving schemes that have low complexities for
this relaying problem. 

The organization of the material in this paper is as follows.
In Section II we describe the deterministic model of relay networks
and the model of the network that we study in this paper. Furthermore,
we discuss the general linear coding and decoding strategies and the
achievable rate of the linear schemes. In Section III we find optimal
linear coding schemes and their corresponding achievable rates.

\section{Problem Setting}

First we briefly state the linear deterministic model of
relay networks from \cite{avestimehr_1,avestimehr_2}, then we introduce our considered network model.

\subsection{Deterministic model}

Consider a directed graph $N(V,E)$ where $V$ denotes the set of
nodes in the network including source, relays and destination and
$E$ is the set of edges. Communication from node $i$ to $j$ has
a nonnegative gain $n_{i,j}$ associated with it. This number models
the channel gain in the corresponding Gaussian setting. Each node
$i$ transmits a vector $x_{i}\in\mathbb{F}_{2}^{q}$ and receives
a vector $y_{i}\in\mathbb{F}_{2}^{q}$ where $q=\max_{i,j}(n_{(i,j)})$.
The received signal at each node is a deterministic function of transmitted
signals at the other nodes with the following input--output relation:
\begin{equation}
y_{j}=\sum_{k:(k,j)\in E}Q^{q-n_{(k,j)}}x_{k}\label{eq:linear_relation}\end{equation}
where $Q$ is the $q\times q$ shift matrix given by\[
Q=\left(\begin{array}{ccccc}
0 & 0 & 0 & \cdots & 0\\
1 & 0 & 0 & \cdots & 0\\
0 & 1 & 0 & \cdots & 0\\
\vdots & \vdots & \ddots & \vdots & \vdots\\
0 & 0 & \cdots & 1 & 0\end{array}\right).\]
In this paper, we are interested in linear coding schemes where at
every node $j$ the transmitted signal is a linear function of the
received signal

\begin{equation}
x_{j}=G_{j}y_{j}.\label{eq:linear coding shceme}\end{equation}
The goal is to design coding functions $G_{i}$ such that the destination
node receives enough information for decoding the message sent by
the source.

\subsection{Diamond network with a disturbing node}

\begin{figure}
\centering\includegraphics[scale=0.5]{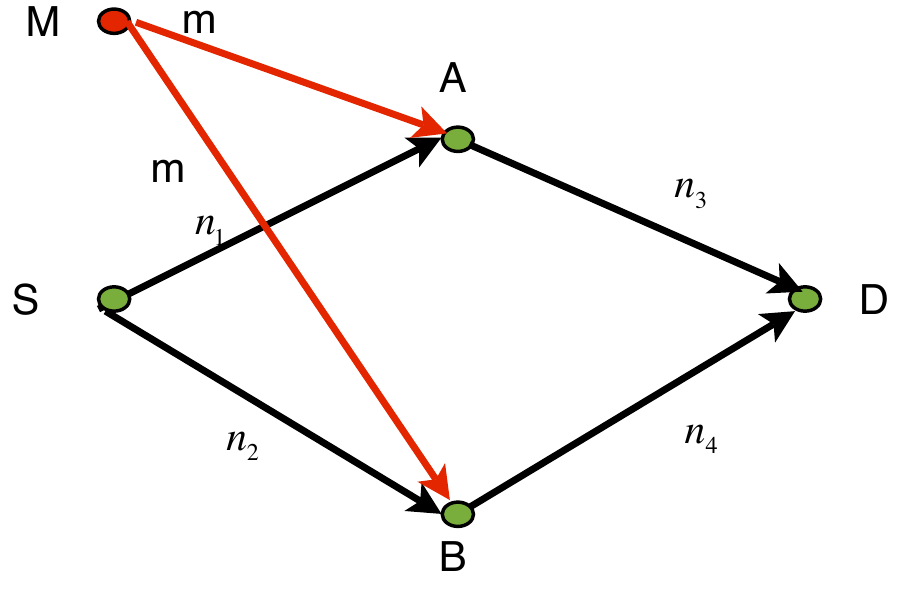}\caption{\label{fig:Diamond-network-with}Diamond network with a disturbing
node}

\end{figure}

We consider a diamond relay network with a disturbing node which is
depicted in Figure \ref{fig:Diamond-network-with}. Here $S$ and
$D$ are respectively the source and the destination nodes and $M$
is the source of disturbing signals. Also nodes $A$ and $B$ are
the relay nodes. For simplicity of our analysis we assume that the
links from $M$ to the two relays have the same gain $m$ that is
also realistic in the situation where the distance between the nodes
in the diamond wireless network is relatively small compared to the
distance from node $M$ which might be operating in a different network.
As before, we denote the transmitted signal from node $i$ by $x_{i}$
and the received signal by $y_{i}.$ 

Let $G_{A}$ and $G_{B}$ be the coding matrices at nodes $A$ and
$B$ respectively. The signal $y_{D}$ received by destination is
a linear combination of the transmitted signals $x_{S}$ and $x_{M}$.
Let \[
y_{D}=G_{S}x_{S}+G_{M}x_{M}.\]
 where by application of (\ref{eq:linear_relation}) we have \begin{align}
G_{S} & =Q^{q-n_{3}}G_{A}Q^{q-n_{1}}+Q^{q-n_{4}}G_{B}Q^{q-n_{2}},\label{eq:G_S}\\
G_{M} & =Q^{q-n_{3}}G_{A}Q^{q-m}+Q^{q-n_{4}}G_{B}Q^{q-m}.\label{eq:G_M}\end{align}
 For every choice of $G_{A}$ and $G_{B}$ we let $R=R(G_{A},G_{B})$
denote the rate of transmission of information from $S$ to $D.$
Also we let $C=\max_{G_{A},G_{B}}R(G_{A},G_{B})$ denote the capacity
of transmission from $S$ to $D$. In the following, we explicitly
find $R(G_{A},G_{B})$ and in the next section we find the capacity
$C$ in terms of the network parameters.

In our analysis, we usually work with the range and dimension of matrices.
For matrix $H$ we let $\mbox{range}(H)$ denote the linear span of
the columns of $H$. Also for a subspace $\mathcal{S}\subseteq\mathbb{F}_{2}^{q}$
we let $\mbox{dim}(\mathcal{S})$ denote its dimension. Obviousely
$\mbox{dim}(\mbox{range}(H))=\mbox{rank}(H).$ For two matrices $H_{1},H_{2}$
we also use the shorthand $\mbox{rank}(H_{1}\cap H_{2})$ to denote
$\mbox{dim}(\mbox{range}(H_{1})\cap\mbox{range}(H_{2})).$
\begin{thm}
\label{thm:rate}For any choice of $G_{A}$ and $G_{B}$ we have\[
R(G_{A},G_{B})=\mbox{rank}(G_{S})-\mbox{rank}(G_{S}\cap G_{M})\]
where $G_{S}$ and $G_{M}$ are defined in (\ref{eq:G_S}) and (\ref{eq:G_M}).
\end{thm}
For two subspaces $\mathcal{S}_{1}$ and $\mathcal{S}_{2}$ of $\mathbb{F}_{2}^{q}$,
let $\mathcal{S}_{1}+\mathcal{S}_{2}=\left\{ s_{1}+s_{2}:s_{1}\in\mathcal{S}_{1},s_{2}\in\mathcal{S}_{2}\right\} $.
To prove the Theorem \ref{thm:rate} we first prove the following
lemma.
\begin{lem}
\label{lem:subspaces}If $\mathcal{S}_{1}$ and $\mathcal{S}_{2}$
are two subspaces of $\mathbb{F}_{2}^{q}$ then for any $s\in\mathcal{S}=\mathcal{S}_{1}+\mathcal{S}_{2}$
there exists a unique pair $s_{1}\in\mathcal{S}_{1}$ and $s_{2}\in\mathcal{S}_{2}$
with $s_{1}+s_{2}=s$ if and only if $\mathcal{S}_{1}\cap\mathcal{S}_{2}=\left\{ \mathbf{0}\right\} .$ \end{lem}
\begin{proof}
First suppose that $\mathcal{S}_{1}\cap\mathcal{S}_{2}\neq\left\{ \mathbf{0}\right\} $
hence, there exists at last $t\neq\mathbf{0}$ such that $t\in\mathcal{S}_{1},t\in\mathcal{S}_{2}.$
Then if $s=s_{1}+s_{2}$ with $s_{1}\in\mathcal{S}_{1}$ and $s_{2}\in\mathcal{S}_{2}$,
we also have $s_{1}+t\in\mathcal{S}_{1}$ and $s_{2}-t\in\mathcal{S}_{2}$
and $(s_{1}+t)+(s_{2}-t)=s$. Therefore the condition is necessary.
To prove the sufficiency, we notice that $\mathcal{S}$ is a subspace
of $\mathbb{F}_{2}^{q}$ and if $\mathcal{S}_{1}\cap\mathcal{S}_{2}=\left\{ \mathbf{0}\right\} ,$
we can form a basis for $\mathcal{S}$ by union of a basis of $\mathcal{S}_{1}$
and a basis of $\mathcal{S}_{2}.$ Now if $s=s_{1}+s_{2}=t_{1}+t_{2}$
and $s_{1}\neq t_{1},s_{2}\neq t_{2}$ we can form two different expansions
for $s$ in the basis of $\mathcal{S}$ by using either the expansion
of $s_{1}$and $s_{2}$ in the bases of $\mathcal{S}_{1}$ and $\mathcal{S}_{2}$
or the expansions of $t_{1}$ and $t_{2}$ in the bases of $\mathcal{S}_{1}$
and $\mathcal{S}_{2}$. This contradicts the fact that each $s\in\mathcal{S}$
has a unique expansion in the basis of $\mathcal{S}$.
\end{proof}
Next we prove Theorem \ref{thm:rate}.
\begin{proof}
For a linear scheme we choose a subspace $\mathcal{X}\subseteq\mathbb{F}_{2}^{q}$
as the set of our codewords $x_{S}$. Let $\mathcal{S}=\left\{ G_{S}x_{S}:x_{S}\in\mathcal{X}\right\} .$
For a successful decoding $\mathcal{X}$ has to satisfy two properties;
first, each $x_{S}\in\mathcal{X}$ should be mapped into a unique
vector in $\mathcal{S}$. This implies that $\mbox{dim}(\mathcal{S})=\mbox{dim}(\mathcal{X}).$
Second, for every $x_{S}\in\mathcal{X},$ and every $x_{M}\in\mathbb{F}_{2}^{q},$
$y_{D}=G_{S}x_{S}+G_{M}x_{M}$ corresponds to the unique pair of $G_{S}x_{S}$
and $G_{M}x_{M}.$ These two conditions guarantee that each $y_{D}$
corresponds to a unique codeword $x_{S}.$ Then, the maximum dimension
of $\mathcal{X}$ that satisfies the two conditions is the rate $R$
of the code. Since $G_{M}x_{M}$ can be any vector in $\mbox{range}(G_{M}),$
to satisfy the second condition, by Lemma \ref{lem:subspaces}, $\mbox{range}(G_{M})\cap\mathcal{S}=\left\{ \mathbf{0}\right\} .$
$\mathcal{S}$ has the maximum dimension when it is the largest subspace
of the set $(\mbox{range}(G_{S})-\mbox{range}(G_{M}))\cup\left\{ \mathbf{0}\right\} $
which has a dimension of $\mbox{rank}(G_{S})-\mbox{rank}(G_{S}\cap G_{M}).$
Next we choose $\mathcal{X}$ to be the subspace of $\mathbb{F}_{2}^{q}$
with dimension of $\mbox{rank}(G_{S})-\mbox{rank}(G_{S}\cap G_{M})$
such that $G_{S}$ maps it to the set $\mathcal{S}.$ Notice that
since $\mathcal{S}$ is a subspace of $\mbox{range}(G_{S})$ we can
always find such $\mathcal{X}.$ Therefore $R=\mbox{dim}(\mathcal{X})=\mbox{dim}(\mathcal{S})=\mbox{rank}(G_{S})-\mbox{rank}(G_{S}\cap G_{M}).$
\end{proof}

\section{Capacity of the Network}

In this section, we find the linear capacity of the network, that is
the maximum achievable rate by a linear coding scheme. In several
steps of our capacity calculation we will use the following useful
lemma:
\begin{lem}
\label{lem:F_and_G}Let $F_{m\times n}$ and $G_{m\times n}$ be two
matrices that are the same in at least their first $m-\alpha$ rows.
Then
\end{lem}
\[
\mbox{rank}(F)-\mbox{rank}(F\cap G)\leq\min(n,\alpha).\]

\begin{proof}
Let $F=\left[\begin{array}{c}
A\\
B\end{array}\right]$ and $G=\left[\begin{array}{c}
A\\
D\end{array}\right]$ where $A$ is a $(m-\alpha)\times n$ matrix. It is easy to verify
that \[
\mbox{rank}(F)-\mbox{rank}(F\cap G)=\mbox{rank}(\left[\begin{array}{cc}
F & G\end{array}\right])-\mbox{rank}(G).\]
 We have \begin{align*}
\mbox{rank}(\left[\begin{array}{cc}
F & G\end{array}\right]) & =\mbox{rank}(\left[\begin{array}{cc}
A & A\\
B & D\end{array}\right])\leq\\
\mbox{rank}(\left[A\right])+\mbox{rank}(\left[\begin{array}{cc}
B & D\end{array}\right]) & \leq\mbox{rank}(A)+\alpha.\end{align*}
 Also, $\mbox{rank}(G)=\mbox{rank}(\left[\begin{array}{c}
A\\
D\end{array}\right])\geq\mbox{rank}(A).$ Therefore $\mbox{rank}(\left[\begin{array}{cc}
F & G\end{array}\right])-\mbox{rank}(G)\leq\alpha$. Also, $\mbox{rank}(F)\leq n$ and thus $\mbox{rank}(F)-\mbox{rank}(F\cap G)\leq n.$
\end{proof}
Next we derive the linear capacity of the network in Figure \ref{fig:Diamond-network-with}.
By symmetry, we only derive the capacity for $n_{1}\geq n_{2}.$

\subsubsection{$n_{1}>n_{2}$, $n_{3}\geq n_{4}$ and $m\leq n_{1}$ }

Let $A$ be a $q\times q$ matrix. For two integers $n_{l}$ and $n_{r}$
consider a partition of the elements of $A$ as follows \[A=\bordermatrix{& q-n_r & n_r \cr                 
n_l & * &  A_1  \cr                 
q-n_l & *  &  * \cr                 
}\]then\[Q^{q-n_{l}}AQ^{q-n_{r}}=\bordermatrix{ & n_{r} & q-n_{r} \cr 
q-n_{l} &  \mathbf{0} & \mathbf{0}\cr 
n_{l} & A_{1} & \mathbf{0} \cr }.\]Now let $G_{A}$ be partitioned as follows \[G_A=\bordermatrix{& q-m & m \cr                 
n_3-n_4 & * &  A  \cr                 
q-n_3+n_4 & *  &  * \cr                 
}\]Then \[Q^{q-n_{3}}G_{A}Q^{q-n_{1}}=
\bordermatrix{& n_1-m & m &q-n_1\cr                 
q-n_3 & {\mathbf 0} &  {\mathbf 0}& {\mathbf 0}  \cr                 
n_3-n_4 & *  &  A & {\mathbf 0} \cr
n_4 &* &* & {\mathbf 0} \cr                
},\]
\[Q^{q-n_{4}}G_{B}Q^{q-n_{2}}=
\bordermatrix{& n_1-m & m &q-n_1\cr                 
q-n_3 & {\mathbf 0} &  {\mathbf 0}& {\mathbf 0}  \cr                 
n_3-n_4 & *  &  {\mathbf 0} & {\mathbf 0} \cr
n_4 &* &* & {\mathbf 0} \cr                
}.\]Therefore, \[G_S=
\bordermatrix{& n_1-m & m &q-n_1\cr                 
q-n_3 & {\mathbf 0} &  {\mathbf 0}& {\mathbf 0}  \cr                 
n_3-n_4 & B_1  &  A & {\mathbf 0} \cr
n_4 &B_2 & D & {\mathbf 0} \cr                
}.\]Also\[Q^{q-n_{3}}G_{A}Q^{q-m}=
\bordermatrix{& m & q-m \cr                 
q-n_3 & {\mathbf 0} &  {\mathbf 0}  \cr                 
n_3-n_4 & A  &   {\mathbf 0} \cr
n_4 &*  & {\mathbf 0} \cr                
},\]
\[Q^{q-n_{4}}G_{B}Q^{q-m}=
\bordermatrix{& m & q-m \cr                 
q-n_3 & {\mathbf 0} &  {\mathbf 0}  \cr                 
n_3-n_4 & {\mathbf 0}  &   {\mathbf 0} \cr
n_4 &*  & {\mathbf 0} \cr                
}.\]Therefore,\[G_M=
\bordermatrix{& m & q-m \cr                 
q-n_3 & {\mathbf 0} &  {\mathbf 0}  \cr                 
n_3-n_4 & A  &   {\mathbf 0} \cr
n_4 &E  & {\mathbf 0} \cr                
}.\]

Next we find an upper bound on any acheivable rate $R=\mbox{rank}(G_{S})-\mbox{rank}(G_{S}\bigcap G_{M})$
in terms of the network parameters and then find $G_{A}$ and $G_{B}$
that achieve the bound. 

\begin{align*}
R & =\mbox{rank}(G_{S})-\mbox{rank}(G_{S}\cap G_{M})\\
 & \leq\mbox{rank}(\left[\begin{array}{c}
B_{1}\\
B_{2}\end{array}\right])+\mbox{rank}(\left[\begin{array}{c}
A\\
D\end{array}\right])-\mbox{rank}(G_{S}\cap G_{M})\\
 & \leq\mbox{rank}(\left[\begin{array}{c}
B_{1}\\
B_{2}\end{array}\right])+\left(\mbox{rank}(\left[\begin{array}{c}
A\\
D\end{array}\right])-\mbox{rank}(\left[\begin{array}{c}
A\\
D\end{array}\right]\cap\left[\begin{array}{c}
A\\
E\end{array}\right])\right)\\
 & \overset{(a)}{\leq}n_{1}-m+\min(m,n_{4}),\end{align*}
 where $(a)$ follows by Lemma \ref{lem:F_and_G}. On the other hand,
$R$ can not exceed the value of any cut set in the absence of the
noisy source, hence\[
R\leq\max(n_{3},n_{4})=n_{3}\]
if we let $k=\min(m,n_{4})$ then we have \begin{equation}
R\leq\min(n_{1}-m+k,n_{3}).\label{eq:bound1}\end{equation}
 We let matrices $G_{A}$ and $G_{B}$ be of the following forms and
show that they achieve the above upper bound. Let $j=\min(n_{1}-m,n_{3}-k)$
and for any $t$ let $I_{t}$ be the $t\times t$ identity matrix.\[G_A=\bordermatrix {&q-n_1 &j&n_1-j-k&  k\cr
j       & \mathbf 0 & I_j       & \mathbf  0 & \mathbf  0 \cr
n_3-j-k & \mathbf 0 & \mathbf 0 & \mathbf 0  & \mathbf 0  \cr
k       & \mathbf 0 & \mathbf 0 & \mathbf 0  &  I_k   \cr
q-n_3   & \mathbf 0 & \mathbf 0 & \mathbf 0  & \mathbf 0 \cr }, \]
\[
G_B=
\bordermatrix{& q-k & k \cr                 
n_4-k & {\mathbf 0} &  {\mathbf 0}  \cr                 
k & \mathbf 0  &   I_k \cr
q- n_4 &\mathbf 0  & {\mathbf 0} \cr                
}.\]Then it is easy to verify that $G_{S}$ is of the following form \[G_S=\bordermatrix {&j&n_1-j-k&  k &q-n_1\cr
q-n_3   & \mathbf 0 & \mathbf 0 & \mathbf 0  & \mathbf 0 \cr
j     & I_j  & \mathbf 0  & \mathbf  0 & \mathbf  0 \cr
n_3-j-k & \mathbf 0 & \mathbf 0 & \mathbf 0  & \mathbf 0  \cr
k       & *  & *   & L    & \mathbf 0     \cr },
\] where $L$ is a lower triangular matrix with ones on its diagonal.
This implies that $\mbox{rank}(G_{S})=j+k.$ Also $G_{M}=\mathbf{0}_{q\times q}$
and hence a rate of $R=j+k$ is achievable by this coding scheme.
This matches our bound ($\ref{eq:bound1}$).

\subsubsection{$n_{1}>n_{2}$, $n_{3}\geq n_{4}$ and $m\geq n_{1}$}

In this case $G_{S}$ and $G_{M}$ are as the following formats:\[G_S=
\bordermatrix{& n_1 & q-n_1 \cr                 
q-n_3 & {\mathbf 0} &  {\mathbf 0}  \cr                 
n_3-n_4 & A  &   {\mathbf 0} \cr
n_4 &D  & {\mathbf 0} \cr                
}\]
\[G_M=
\bordermatrix{& m-n_1 & n_1 &q-m\cr                 
q-n_3 & {\mathbf 0} &  {\mathbf 0}& {\mathbf 0}  \cr                 
n_3-n_4 & C_1  &  A & {\mathbf 0} \cr
n_4 &C_2 & E & {\mathbf 0} \cr                
}\] We have the following upper bound on any achievable rate $R:$\begin{align}
R & =\mbox{rank}(G_{S})-\mbox{rank}(G_{S}\cap G_{M})\nonumber \\
 & \leq\mbox{rank}(\left[\begin{array}{c}
A\\
D\end{array}\right])-\mbox{rank}(\left[\begin{array}{c}
A\\
D\end{array}\right]\cap\left[\begin{array}{c}
A\\
E\end{array}\right])\nonumber \\
 & \overset{(a)}{\leq}\min(n_{1},n_{4}),\label{eq:bound2}\end{align}
 where $(a)$ follows by Lemma \ref{lem:F_and_G}. Let $k=\min(n_{1},n_{4}).$
We design $G_{A}$ and $G_{B}$ as follows:\[G_A= \bordermatrix {& q-n_1 &n_1-k &k \cr 
n_3-k & \mathbf 0 & \mathbf 0 & \mathbf 0 \cr
k & \mathbf 0 & \mathbf 0  & I_k \cr
q-n_3 & \mathbf 0 & \mathbf 0  & \mathbf  0 \cr},\]  
\[G_B= \bordermatrix {& q-n_1 &n_1-k &k \cr 
n_4-k & \mathbf 0 & \mathbf 0 & \mathbf 0 \cr
k & \mathbf 0 & \mathbf 0  & I_k \cr
q-n_4 & \mathbf 0 & \mathbf 0  & \mathbf  0 \cr}.\]Then $G_{M}=\mathbf{0}_{q\times q}$ and $G_{S}$ is of the following
form\[G_S= \bordermatrix {& n_1-k & k & q-n_1  \cr
q-k & \mathbf 0 & \mathbf 0 & \mathbf0  \cr 
k & * & L & \mathbf 0} 
,\] where $L$ is a lower triangular matrix with ones on its diagonal.
Therefore $\mbox{rank}(G_{S})=k$ and $R=k$ is achievable. This matches
bound ($\ref{eq:bound2}$).

\subsubsection{$n_{1}>n_{2}$, $n_{4}\geq n_{3}$ and $m\leq n_{2}$}

In this case for any choices of $G_{A}$ and $G_{B}$,$G_{S}$ and
$G_{M}$ are of the following forms:\[G_S= \bordermatrix{& n_2-m & m & n_1-n_2 & q-n_1 \cr
q-n_4 & \mathbf 0 & \mathbf 0 & \mathbf 0 & \mathbf 0 \cr
n_4-n_3 & B_1 & A & \mathbf 0 & \mathbf 0 \cr
n_3 & B_2 & D_1 & D_2 & \mathbf 0 \cr
},
\]
\[G_M= \bordermatrix{ & m & q-m \cr
q-n_4 & \mathbf 0 & \mathbf 0 \cr
n_4-n_3 & A & \mathbf 0 \cr
n_3 & E & \mathbf 0 \cr
}.
\]Let us define\[\hat{G}_{M} = \bordermatrix{&m & n_1-n_2 \cr
n_4-n_3 & A & \mathbf 0 \cr
n_3 & E & \mathbf 0 \cr}.
\]The following bound holds for any rate $R:$ \begin{align*}
R & =\mbox{rank}(G_{S})-\mbox{rank}(G_{S}\cap G_{M})\\
 & \leq\mbox{rank}(\left[\begin{array}{c}
B_{1}\\
B_{2}\end{array}\right])+\mbox{rank}(\left[\begin{array}{cc}
A & \mathbf{0}\\
D_{1} & D_{2}\end{array}\right])\\
 & \quad-\mbox{rank}(\left[\begin{array}{cc}
A & \mathbf{0}\\
D_{1} & D_{2}\end{array}\right]\cap\hat{G}_{M})\\
 & \overset{(a)}{\leq}n_{2}-m+n_{3},\end{align*}
where $(a)$ follows by Lemma \ref{lem:F_and_G}. This bound, together
with the cutset bound $R\leq\min(n_{1},n_{4},n_{2}+n_{3})$ in the
absence of the node $M,$ results in the following bound\begin{equation}
R\leq\min(n_{1},n_{4},n_{2}+n_{3}-m).\label{eq:bound3}\end{equation}

Let $k=\min(n_{3},m)$. We design $G_{A}$ and $G_{B}$ as follows:\[G_A=\bordermatrix {&q-n_1 &n_2-m&n_1-n_2&  m-k&k\cr
n_3-k       & \mathbf 0 & F_1       & F_2   & \mathbf  0 &\mathbf  0 \cr
k & \mathbf 0 & \mathbf 0 & \mathbf 0  & \mathbf 0 &I_k \cr
q-n_3   & \mathbf 0 & \mathbf 0 & \mathbf 0  & \mathbf 0 &\mathbf  0 \cr }, \]
\[
G_B=
\bordermatrix{& q-n_2 & n_2-m &m-k &k \cr                 
n_4-n_3 & \mathbf 0 & F_3 & \mathbf 0 & \mathbf 0  \cr                 
n_3-k & \mathbf 0  & \mathbf 0 & \mathbf 0 & \mathbf 0 \cr
k & \mathbf 0  & \mathbf 0 & \mathbf 0 & I_k \cr
q- n_4 &\mathbf 0  & \mathbf 0 & \mathbf 0 & \mathbf 0 \cr             
}\]where $F_{1}$, $F_{2}$ and $F_{3}$ are chosen so that $\mbox{rank}(F)$
is maximum\[F = \bordermatrix{&n_2-m & n_1-n_2 \cr
n_4-n_3 & F_3 & \mathbf 0 \cr
n_3-k & F_1 & F_2 \cr}.
\]We know that: \[
\max(\mbox{rank}(F))=\min(n_{4}-k,n_{1}-m,n_{2}+n_{3}-m-k).\]
By this choice, $G_{M}=\mathbf{0}_{q\times q}$ and $G_{S}$ is of
the following form\[G_S=
\bordermatrix{& n_2-m & n_1-n_2 & m-k & k & q-n_1 \cr                 
q-n_4 & {\mathbf 0} &  {\mathbf 0} & {\mathbf 0} & {\mathbf 0} & {\mathbf 0}  \cr                 
n_4-n_3 & F_3  & {\mathbf 0} & {\mathbf 0} & {\mathbf 0} & {\mathbf 0} \cr
n_3-k & F_1  & F_2 & {\mathbf 0} & {\mathbf 0} & {\mathbf 0} \cr
k & {\mathbf 0} & * & * & L & {\mathbf 0} \cr               
},\]where $L$ is a lower triangular matrix with ones on its diagonal.\\
Therefore \begin{align*}
\mbox{rank}(G_{S})= & k+\mbox{rank}(F)\\
= & \min(n_{4}-k,n_{1}-m,n_{2}+n_{3}-m-k)+k\\
= & \min(n_{4},n_{1}-m+k,n_{2}+n_{3}-m)\\
= & \min(n_{4},n_{1}-m+n_{3},n_{1}-m+m,n_{2}+n_{3}-m)\\
= & \min(n_{1},n_{4},n_{2}+n_{3}-m)\end{align*}
and $R=\min(n_{1},n_{4},n_{2}+n_{3}-m)$ is achievable. This matches
our bound ($\ref{eq:bound3}$).

\subsubsection{$n_{1}>n_{2}$, $n_{4}\geq n_{3}$ and $m>n_{2}$}

In this case $G_{S}$ and $G_{M}$ are of the following forms\[G_S=\bordermatrix {& n_2 &n_1-n_2 &q-n_1 \cr
q-n_4 & \mathbf 0 & \mathbf 0 & \mathbf 0 \cr
n_4-n_3 & A & \mathbf 0 & \mathbf 0 \cr
n_3 & D_1 & D_2 & \mathbf 0 \cr
},\]
\[G_M=\bordermatrix{& m-n_2 &n_2 & q-m \cr
q-n_4 & \mathbf 0 & \mathbf 0 & \mathbf 0 \cr
n_4-n_3 & C_1 & A & \mathbf 0 \cr
n_3 & C_2 & E & \mathbf 0 \cr
}.
\]Let us define the following matrix\[\hat{G}_{M}=\bordermatrix {& n_2 & n_1-n_2 \cr
q-n_4 & \mathbf 0 & \mathbf 0 &  \cr
n_4-n_3 & A & \mathbf 0 \cr
n_3 & E & \mathbf 0 \cr
}.
\]Then we have the following bound on $R$:\begin{align}
R & =\mbox{rank}(G_{S})-\mbox{rank}(G_{S}\cap G_{M})\nonumber \\
 & \leq\mbox{rank}(G_{S})-\mbox{rank}(G_{S}\cap\hat{G}_{M})\nonumber \\
 & \overset{(a)}{\leq}\min(n_{1},n_{3}),\label{eq:bound4}\end{align}
where $(a)$ follows by Lemma \ref{lem:F_and_G}. Let $k=\min(n_{1},n_{3}).$
We design $G_{A}$ and $G_{B}$ as follows:\[G_A=\bordermatrix{& q-k & k \cr
n_3-k & \mathbf 0 & \mathbf 0 \cr
k & \mathbf 0 & I_k \cr
q-n_3 & \mathbf 0 & \mathbf 0 \cr
},\]
\[G_B= \bordermatrix{& q-k & k \cr
n_4-k & \mathbf 0 & \mathbf 0 \cr
k & \mathbf 0 & I_k \cr
q-n_4 & \mathbf 0 & \mathbf 0 \cr
}.
\]It is easy to verify that $G_{M}=\mathbf{0}_{k\times k}$ and $G_{S}$
is of the following form\[G_S=\bordermatrix{& n_1-k & k & q-n_1 \cr
q-k & \mathbf 0 & \mathbf 0 & \mathbf 0 \cr
k & * & L & \mathbf 0 \cr
},
\]where $L$ is a lower triangular matrix with ones on its diagonal.
Therefore $R=\mbox{rank}(G_{S})=k.$ This matches the bound ($\ref{eq:bound4}$).

\subsubsection{$n_{1}=n_{2}$ , $m\geq n_{1}$ }

In this case it is easy to verify that for any choice of $G_{A}$
and $G_{B}$, columns of $G_{S}$ are a subset of columns of $G_{M}$.
Therefore $\mbox{rank}(G_{S}\cap G_{M})=\mbox{rank}(G_{S})$ and hence
$R=\mbox{rank}(G_{S})-\mbox{rank}(G_{S}\cap G_{M})=0.$

\subsubsection{$n_{1}=n_{2}$ , $m<n_{1}$}

In this case for any choice of $G_{A}$ and $G_{B}$, columns of $G_{M}$
are a subset of columns of $G_{S}$. Therefore $\mbox{rank}(G_{S}\cap G_{M})=\mbox{rank}(G_{M}),$
and $R=\mbox{rank}(G_{S})-\mbox{rank}(G_{M}).$ Since there are at
most $n_{1}-m$ columns in $G_{S}$ that does not appear in $G_{M}$,
$\mbox{rank}(G_{S})-\mbox{rank}(G_{M})\leq n_{1}-m$ and therefore
$R\leq n_{1}-m$. Also from the cutset bound $R\leq\max(n_{3},n_{4})$
we have \begin{equation}
R\leq\min(n_{1}-m,\max(n_{3},n_{4})).\label{eq:bound5}\end{equation}
 Let $k=\min(n_{1}-m,\max(n_{3},n_{4}))$. Consider two cases:
\begin{itemize}
\item If $n_{3}\geq n_{4}$ then we set $G_{B}=\mathbf{0}_{q\times q}$
and let \[G_A=\bordermatrix{& q-n_1 & k& n_1-k \cr
k & \mathbf 0 & I_k & \mathbf 0 \cr
q-k & \mathbf 0 & \mathbf 0 & \mathbf 0 \cr}.
\]In this case $G_{M}=\mathbf{0}_{q\times q}$ and \[G_S=\bordermatrix{& k & q-k \cr
q-n_3 & \mathbf 0 & \mathbf 0 \cr
k & I_k & \mathbf 0 \cr
n_3-k & \mathbf 0 & \mathbf 0 \cr
}.
\] Therefore $R=\mbox{rank}(G_{S})-\mbox{rank}(G_{M})=k$, that achieves
bound $\eqref{eq:bound5}$.
\item If $n_{4}>n_{3}$ then we set $G_{A}=\mathbf{0}_{q\times q}$ and
let\[G_B=\bordermatrix{& q-n_1 & k & n_1-k \cr
k & \mathbf 0 & I_k & \mathbf 0 \cr
q-k & \mathbf 0 & \mathbf 0 & \mathbf 0 \cr}.
\]Similar to the previous case, in this case $G_{M}=\mathbf{0}_{q\times q}$
and \[G_S=\bordermatrix{& k & q-k \cr
q-n_4 & \mathbf 0 & \mathbf 0 \cr
k & I_k & \mathbf 0 \cr
n_4-k & \mathbf 0 & \mathbf 0 \cr
}.
\]Therefore $R=\mbox{rank}(G_{S})-\mbox{rank}(G_{M})=k$, that achieves
bound $\eqref{eq:bound5}$.\end{itemize}
\begin{rem}
It is easy to verify that for cases 1, 3 and 6 where $m$ can be set to zero, the capacity of diamond network   \cite{avestimehr_1} is achievable by our coding scheme. \end{rem}
\section*{Acknowledgment} 
The authors wish to thank  Bahareh Akhbari for her comments.

\end{document}